\documentclass[epj]{svjour}
\usepackage{graphics}
\usepackage{epsfig}
\def\la{\;\raise0.3ex\hbox{$<$\kern-0.75em\raise-1.1ex\hbox{$\sim$}}\;}
\def\ga{\;\raise0.3ex\hbox{$>$\kern-0.75em\raise-1.1ex\hbox{$\sim$}}\;}
\def\lr{\;\raise0.3ex\hbox{$\rightarrow$\kern-1.0em\raise-1.1ex\hbox{$\leftarrow$}}\;}

\begin{document}
\title{Pycnonuclear reactions in dense stellar matter}

\author{D.G.\ Yakovlev \inst{1} \and K.P.\ Levenfish \inst{1}
\and O.Y.\ Gnedin \inst{2} % etc
}                     % Do not remove
\offprints{}          % Insert a name or remove this line
\institute{Ioffe Physical Technical Institute, Politekhnicheskaya 26,
         St.-Petersburg, 194021, Russia
   \and 
   Ohio State University,
        760 1/2 Park Street, Columbus, OH 43215, USA}
\date{Received: date / Revised version: date}
\abstract{
We discuss pycnonuclear burning of
highly exotic atomic nuclei in deep crusts of neutron stars,
at densities up to $10^{13}$ g cm$^{-3}$.
As an application, we consider pycnonuclear burning of matter accreted
on a neutron star in a soft X-ray transient (SXT, a compact binary
containing a neutron star and a low-mass companion). 
The energy released in this burning,
while the matter sinks into the stellar crust
under the weight of newly accreted material, is sufficient
to warm up the star and initiate neutrino emission
in its core. The surface thermal radiation 
of the star in quiescent states becomes 
dependent of poorly known equation of state (EOS) of supranuclear 
matter in the stellar core, which gives a method to explore
this EOS. Four qualitatively different model EOSs 
are tested
against observations of SXTs. They
imply different levels of the enhancement of neutrino emission
in massive neutron stars by ({\it 1}) the direct Urca process
in nucleon/hyperon matter; ({\it 2}) pion condensates; 
({\it 3}) kaon condensates; ({\it 4}) Cooper pairing of neutrons
in nucleon matter with the forbidden direct Urca process.
A low level of the thermal quiescent emission 
of two SXTs, SAX J1808.4--3658
and Cen X-4, contradicts model ({\it 4}).
Observations of SXTs test the
same physics of dense matter as observations of thermal radiation
from cooling isolated neutron stars, but the data
on SXTs are currently more conclusive.
 % end of PACS codes
} %end of abstract
\maketitle
%
% ######################################################################
%                          TEXT BODY
% ######################################################################

%***************************************************************
\section{Introduction}
\label{sect1}
%***************************************************************

Neutron stars (NSs) are the most compact
stars known in the Universe. 
Their masses are $M \sim 1.4 \, M_\odot$
($M_\odot$ being the solar mass) while
their radii are $R \sim 10$ km. 
They are thought to consist 
(e.g., Ref.\ \cite{lp01,haensel03}) of a 
massive dense core surrounded by a thin crust
(of mass $\la 0.01\,M_\odot$ and thickness $\la 1$ km).
The crust-core interface
occurs at the density 
$\rho \approx \rho_0/2$, where $\rho_0 = 2.8 \times 10^{14}$
g cm$^{-3}$ is the standard density of saturated nuclear matter.

The NS crust is composed of neutron-rich atomic nuclei,
strongly degenerate electrons and (at $\rho \ga 4 \times 10^{11}$
g cm$^{-3}$) of free neutrons dripped from nuclei.  
NS cores contain degenerate matter
of supranuclear density (up to $10-15\, \rho_0$).
Its composition
and equation of state (EOS) are
model dependent (being determined by still
poorly known strong interactions 
in dense matter). 
On the other hand, these properties
are almost not constrained by observations (e.g., Ref.\ \cite{lp01}).
The NS core is usually subdivided into the outer core
($\rho \la 2 \, \rho_0$) and the inner core ($\rho \ga 2 \, \rho_0$).
The outer core is available in all NSs, while the
inner core is present only in massive, more compact NSs.
The outer core is thought to be composed of
neutrons, with the admixture of protons, electrons (and
possibly muons). The composition and EOS of the inner
core is still a mystery. It may be the same composition
as in the outer core, with a possible addition
of hyperons. Alternatively, the inner core
may contain 
pion- or kaon condensates, or quark matter, or the
mixture of these components. Another complication
is introduced by superfluidity of neutrons, protons
and other baryons in dense matter. Superfluid gaps
are very model dependent 
%and cannot be calculated with confidence 
(e.g., Ref.\ \cite{ls01}).

It is unlikely that the fundamental problem of
the EOS of supranuclear matter can be solved on purely
theoretical grounds. However
the solution can be obtained by comparing
theoretical models of NSs with observations. 
We will mention two lines of such studies. The first
one is based on  theory and observations of cooling isolated NSs (e.g.,
Refs.\ \cite{yp04,pageetal04,gusakovetal04,yakovlevetal04}). 
%Theoretical cooling models differ
%mainly because of different neutrino emission properties
%for different EOSs.
%The theory can be compared with observations of thermal
%radiation from isolated NSs.
These studies, carried out over several
decades, have constrained 
the properties of dense matter but
have not solved the problem (Sect.\ 4).
Here, we focus on an alternative method to test
the same physics on another class of objects,
transiently accreting NSs in soft X-ray transients
(SXTs). This method is based on pycnonuclear burning
of accreted matter. 
The method is new but seems to be currently more restrictive.
  
%***************************************************************
\section{Deep crustal heating}
\label{sect2}
%***************************************************************

Nuclear fusion reactions
in normal stars proceed in the {\it thermonuclear}
regime, in which the Coulomb barrier 
is penetrated owing to the thermal energy of colliding nuclei.
Here, we discuss another,
{\it pycnonuclear} regime where the Coulomb barrier
is penetrated due to zero-point vibrations of
nuclei arranged, for instance, in a lattice.
The thermonuclear regime
is realized in a rather low-density and warm plasma,
whereas the pycnonuclear regime operates at high densities and
not too high temperatures. Pycnonuclear reactions
are almost temperature independent and
occur even at $T=0$. They were suggested
by Gamow \cite{gamow39} in 1938. The first calculations
of pycnonuclear reaction rates were done by Wildhack
\cite{wildhack40} upon Gamow's request.
The strict approach for calculating these rates
was formulated by Salpeter and van Horn \cite{svh69} who
studied also
three other regimes of nuclear burning in dense matter
(the thermonuclear regime with strong enhancement due
to plasma screening effects;
the intermediate thermo-pycno nuclear regime;
and the thermally enhanced pycnonuclear regime).
The pycnonuclear regime
has been analyzed later in a number of publications
(e.g., \cite{sc90,kitamura00} and references therein).
Pycnonuclear reactions are extremely slow 
at densities $\rho$ typical for normal stars
but intensify with increasing $\rho$. For example,
carbon burns rapidly into heavier elements at $\rho \ga 10^{10}$
g cm$^{-3}$.

Pycnonuclear
burning can be important in transiently accreting NSs.
%Let us follow the evolution of accreted matter in
%such a star.
The heat released due to the infall of matter and
thermonuclear reactions in the surface layers is radiated away
by photons from the NS surface and cannot
warm up the NS interiors. The accreted matter sinks into
the NS crust under the weight of newly accreted
material. The density gradually
increases in a matter element, and the nuclei undergo
transformations --
beta captures, absorption and
emission of neutrons, and pycnonuclear reactions.
The nuclei evolve then into highly exotic
atomic nuclei which are unstable in laboratory
conditions but stable in dense matter.
The transformations and associated energy release
have been studied by Haensel and Zdunik \cite{hz90,hz03}.
At $\rho \ga 10^9$ g cm$^{-3}$ the transformations are
almost temperature independent but depend on the composition
of matter at $\rho \sim 10^9$ g cm$^{-3}$.
The main energy release
occurs at densities from about $10^{12}$
to $10^{13}$ g cm$^{-3}$, 
several hundred meters under the NS surface, 
in pycnonuclear reactions.
For iron matter at $\rho \sim 10^9$ g cm$^{-3}$,   
these reactions are \cite{hz90}
$^{34}$Ne+$^{34}$Ne$\to ^{68}$Ca;
$^{36}$Ne+$^{36}$Ne$\to ^{72}$Ca; and
$^{48}$Mg+$^{48}$Mg$\to ^{96}$Cr. 
The total energy release is then $\approx$1.45 MeV per
accreted baryon;
the total NS heating power is determined by the mass accretion rate
$\dot{M}$:
\begin{eqnarray}
  L_{\rm dh} & = & 1.45~{\rm MeV} \, \dot{M}/m_{\rm N}
\nonumber \\
  & \approx & 8.74 \times 10^{33}  \,  
  \dot{M}/(10^{-10} \, { M}_\odot \; {\rm yr}^{-1})
  \;\;{\rm   erg\; s}^{-1},
\label{Ldh} 
\end{eqnarray}
where 
$m_{\rm N}$ is the nucleon mass.
It produces {\it deep crustal heating}
of the star.
This heat is spread by thermal conductivity
over the entire NS and warms it up.

%*************************************************************
\section{Thermal states of soft X-ray transients}
\label{sect3}
%*************************************************************

It is possible that the deep crustal heating manifests itself
in NSs which enter SXTs,
compact binaries with low-mass companions \cite{csl97}.
These objects undergo the periods of outburst activity
(days--months, sometimes years)
superimposed with
the periods of quiescence (months--decades).
An active period is thought to be associated with an 
accretion of matter from a companion
through an accretion disk.
The accretion energy released at the
NS surface is so large that a SXT
looks like a bright X-ray source
with the luminosity $L_X \sim 10^{36}-10^{38}$
erg s$^{-1}$. The accretion is
switched off or suppressed in quiescent periods when
$L_X$ drops below $10^{34}$ erg~s$^{-1}$.
In many cases, the spectra of quiescent
emission contain the thermal component, 
well described by NS hydrogen atmosphere models 
with effective surface temperatures from
few $10^5$ K to $\sim 10^6$ K. This can be
the radiation emergent from warm NSs.

The suggestion to interpret the quiescent radiation spectra of SXTs with  
hydrogen atmosphere models was put forward by
Brown et al.\ \cite{bbr98} who assumed also
that NSs in SXTs could be warmed up by the deep crustal heating
of accreted matter.
It is important that the crustal heat is partly radiated away by
neutrinos from the entire NS body. Because the neutrino luminosity 
depends on the NS 
structure, 
the remaining heat, diffused to the surface
and radiated away by photons,
becomes also dependent of the NS structure.
This opens an attractive possibility
(see \cite{ur01,cgpp01,rbbpzu02,bbc02}
and references therein) to explore the NS structure by comparing
the observed quiescent thermal
radiation from SXTs with theoretical predictions.

We outline the results of such studies 
using the theory of thermal states
of transiently accreting NSs described in Ref.\ \cite{ylh03}.
These stars are thermally
inertial, with 
thermal relaxation times 
$\sim10^4$ yr \cite{cgpp01}. In the first approximation, 
a transiently accreting NS is in a (quasi)stationary
steady state 
determined by the mass
accretion rate $\dot{M}\equiv \langle \dot{M} \rangle$
averaged over thermal relaxation time-scales.
Typically, $\langle \dot{M} \rangle$ ranges
from $10^{-14}$ to $10^{-9}$ $M_\odot$ yr$^{-1}$
and does not increase
noticeably the NS mass 
during long periods of NS evolution.
Quasistationary states are expected to be rather insensitive
to variations of the accretion rate associated with
a sequence of active and quiescent states.

%%%%%%%%%%%%%%%%%%%%%%%%%%%%%%%%%%%%%%%%%%%%%%%%%%%%%%%%%%% Fig. 6
\begin{figure}[t]
%\centering
%\begin{center}
\epsfysize=75mm
\epsffile[20 150 460 570]{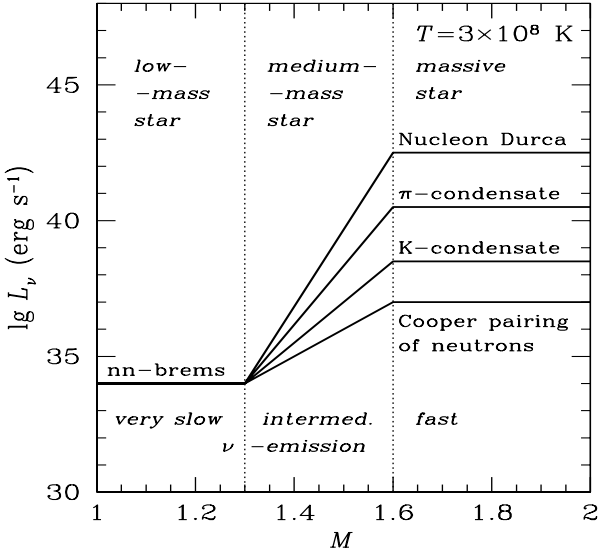}
\caption{ A sketch of the NS neutrino luminosity
$L_\nu$ versus NS mass at the internal stellar
temperature $T=3 \times 10^8$ K for four scenarios of
NS structure (from Ref.\ \cite{yakovlevetal04}). 
}
%\end{center}
\label{four}
\end{figure}
%%%%%%%%%%%%%%%%%%%%%%%%%%%%%%%%%%%%%%%%%%%%%%%%%%%%%%%%%%%%%%%%%%%%%%%%%
 
Thermal states of accreting NSs can be greatly 
affected by the neutrino emission from NS interiors
(mainly from the cores). The neutrino luminosity of
the star, $L_\nu$, in its turn, can strongly depend 
on the NS mass 
(as schematically shown in Fig.\ \ref{four}).
Low-mass NSs possess nucleon cores 
with not too high neutrino emission
determined mainly by the modified Urca (Murca) 
and nucleon-nucleon brems\-strah\-lung (brems) processes.
High-mass NSs, in addition to the outer nucleon cores,
possess the inner cores whose composition 
and EOS are unknown (Sect.\ 1). Their neutrino emission
can be enhanced with respect to the emission
of low-mass NSs. 
Four qualitatively different enhancement mechanisms are
(e.g., Ref.\ \cite{yakovlevetal04}):
({\it 1}) a very powerful direct Urca 
(Durca) process in the cores of massive NSs
containing nucleons (and possibly hyperons); ({\it 2}) 
a less powerful direct-Urca-like process in pion-condensed cores;
({\it 3}) even less powerful similar processes in kaon-condensed
or quark NS cores; 
({\it 4}) even weaker enhancement of the neutrino emission
owing to mild Cooper pairing
of neutrons in the nucleon inner NS cores
with forbidden direct Urca process
\cite{pageetal04,gusakovetal04}.
A transition from
the slow neutrino emission in low-mass NSs to a fast emission
in massive NSs with increasing $M$ occurs in medium-mass NSs;
the mass range of medium-mass stars is model-dependent.
The NS neutrino luminosity is a strong function of the
internal stellar temperature.

A thermal state of the star is determined by
the thermal balance equation 
(which implies that the crustal heating power
is equal to the sum of photon and neutrino luminosities).
Solving this equation, 
one obtains a {\it heating curve}, the dependence of
the thermal photon luminosity (as detected by a distant observer)
on the mean mass accretion rate, $L^\infty_\gamma(\dot{M})$.

%%%%%%%%%%%%%%%%%%%%%%%%%%%%%%%%%%%%%%%%%%%%%%%%%%%%%%%%%% 
\begin{figure}[t]
\centering
\epsfysize=8cm
\epsffile[80 220 550 670]{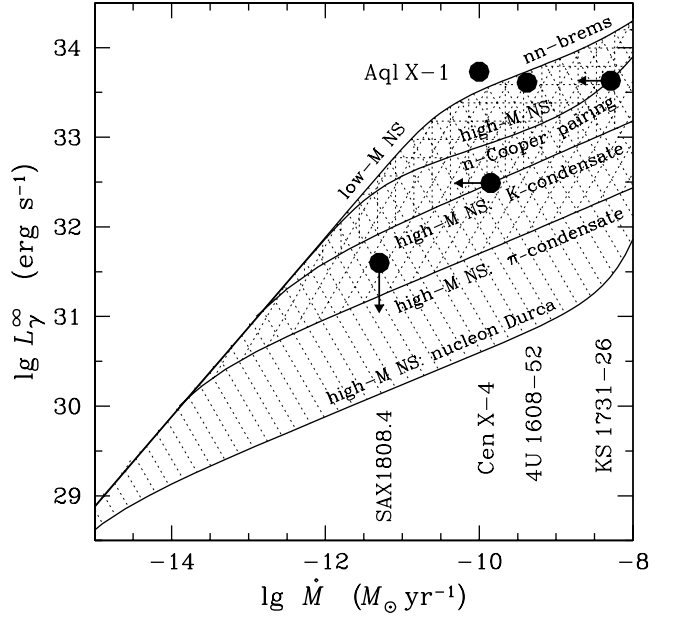}
\caption{\hspace{.2cm}
Quiescent thermal luminosity of several
NSs in SXTs versus mass accretion rate compared with
theoretical heating curves. Four ranges of $L_\gamma^\infty$
(different hatching types) correspond to four scenarios
of neutrino emission (Fig.\ 1). 
}
\label{fig2}
\end{figure}
%%%%%%%%%%%%%%%%%%%%%%%%%%%%%%%%%%%%%%%%%%%%%%%%%%%%%%%%%%%%%%

Several examples are presented in Fig.\ 2.
%The EOS of NS interiors has to be the same
%for all NSs, and we would like to constrain it from
%observations of SXTs. 
The upper curve is calculated for a low-mass NS ($M=1.1\,M_\odot$)
whose core is composed of neutrons, protons
and electrons. 
In the core we use a moderately stiff
EOS proposed in Ref.\ \cite{pal88}
(the same version as employed in Ref.\ \cite{yp04}).
It is assumed that the core contains
strongly superfluid protons. Superfluidity
suppresses the modified Urca process, the star
has a low neutrino luminosity
due to neutron-neutron bremsstrahlung
and stays relatively warm.
The lowest curve in Fig.\ 2 corresponds to the maximum-mass
NS ($M=1.977\, M_\odot$) with the same EOS.
This star has a massive inner core where the direct
Urca process operates and proton superfluidity
dies out. The neutrino luminosity becomes exceptionally
high, and the star is very cold. The next two curves
above the lowest one are schematic models \cite{ylh03}
of high-mass NSs with pion-condensed or kaon-condensed
cores. The next curve is calculated for a high-mass 
($M=2.05\,M_\odot$) star, 
where the direct Urca process is forbidden but the
neutrino emission is enhanced by mild Cooper pairing
of neutrons in the inner core (the same EOS \cite{dh01} and
model for superfluidity as in Fig.\ 1 of Ref.\ 
\cite{gusakovetal04}).

%%%%%%%%%%%%%%%%%%%%%%%%%%%%%%%%%%%%%%%%%%%%%%%%%%%%%%%%%%% Fig. 6
\begin{figure}[t]
%\centering
%\begin{center}
\epsfysize=80mm
\epsffile[20 150 550 670]{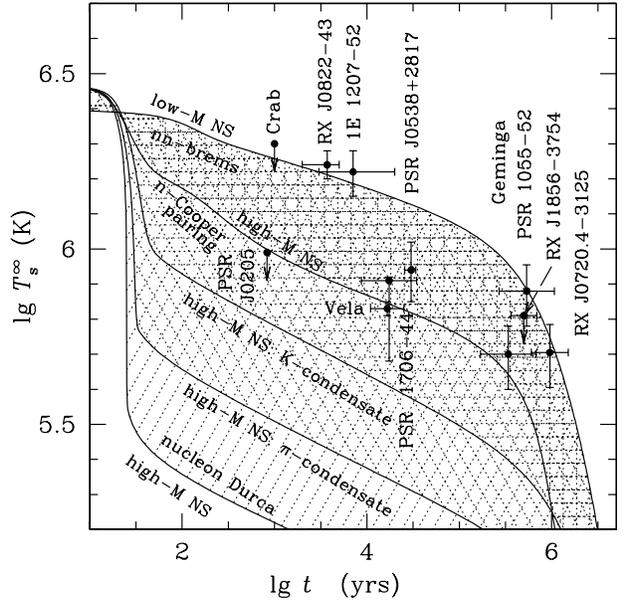}
\caption{ 
Effective surface temperatures $T_{\rm s}^\infty$ of several isolated
NSs versus their ages compared with theoretical cooling curves
(from Ref.\ \cite{yakovlevetal04}).
Four ranges of $T_{\rm s}^\infty$
(different hatching types) correspond to four scenarios
of neutrino emission (Fig.\ 1). 
}
%\end{center}
\label{cool}
\end{figure}
%%%%%%%%%%%%%%%%%%%%%%%%%%%%%%%%%%%%%%%%%%%%%%%%%%%%%%%%%%%%%%%%%%%%%%%%%
 
The heating curve of a low-mass NS
provides an upper 
limit of $L_\gamma^\infty$, 
whereas a heating curve of a high-mass NS
gives a lower limit of $L_\gamma^\infty$,
for a fixed scenario of neutrino emission.
Varying the NS mass from the lower values to the higher
we obtain a family of heating curves which fill in the 
(hatched) space between the upper and lower curves.
In Fig.\ 2 we have four hatched (acceptable) regions for four
scenarios of neutrino emission in Fig.\ 1.

%%%%%%%%%%%%%%%%%%%%%%%%%%%%%%%%%%%%%%%%%
\section{Discussion}
\label{sect4}
%%%%%%%%%%%%%%%%%%%%%%%%%%%%%%%%%%%%%%%%%

These results can be compared with observations of SXTs.
The observational data are mostly the same as in
Ref.\ \cite{ylh03}.
We regard $L_\gamma^\infty$ as the thermal quiescent
luminosity of NSs,
and $\dot{M}$ as the
time-averaged mass accretion rate.
The value of $\dot{M}$ for KS 1731--26
is most probably an upper limit. No quiescent   thermal
emission has been detected \cite{campanaetal02}
from SAX J1808.4--3658,
and we present the upper limit of $L_\gamma^\infty$
for this source. The limit was obtained by P.\ Stykovsky (private
communication, 2004) and discussed
in the note added in proof of Ref.\ \cite{yp04}. The NS seems to be cold,
but the result should be taken with caution because
it is based on one observation (March 24, 2001)
with poor statistics. 
All the data are rather uncertain and are thus plotted
by thick dots.

According to Fig.\ 2,
we can treat NSs in 4U 1608--52 and Aql X-1 as low-mass NSs
(the observations of Aql X-1 
can be explained \cite{coldhot} assuming
the presence of light elements on the NS surface,
that increases $L_\gamma^\infty$).
NSs in Cen X-4 and SAX J1808.4--3658 seem to require the enhanced
neutrino emission and are thus more massive.
The status of the NS in KS 1731--26 is less certain
\cite{ylh03} because
of poorly determined $\dot{M}$.
Similar conclusions have been made 
with respect to some of these sources or selected groups
(see, e.g., \cite{ur01}--\cite{coldhot} and references therein)
although four 
scenarios ({\it 1})--({\it 4}) are discussed 
together for the first time.
%\cite{ur01,cgpp01,rbbpz01,rbbpzu02,rbbpz02,bbc02,wgvm02}.

As seen from Fig.\ 2, the data are still not restrictive.
They cannot allow us to discriminate between
scenarios ({\it 1})--({\it 3}) 
(nucleon NS cores with the direct Urca
process, pion-condensed or kaon condensed cores).
Nevertheless, the observations of Cen X-4 and
especially of SAX J1808.4--3658 seem to contradict
scenario ({\it 4}) of nucleon matter
with the forbidden direct Urca process [and the data on
SAX J1808.4--3658 marginally contradict scenario ({\it 3})].

The same four scenarios can also be tested
by comparing observations of thermal
radiation of cooling (isolated)
NSs with theoretical cooling curves (the dependence
of the effective surface temperatures $T_{\rm s}^\infty$,
redshifted for a distant observer, on the stellar age).
These results are presented in Fig.\ 3; the 
data and curves are taken from
Ref.\ \cite{yakovlevetal04}. 
By comparing Figs.\ 2 and 3 we see
that the data on isolated NSs, although more
numerous, are currently less conclusive and do not discriminate
between all scenarios ({\it 1})--({\it 4}) [although a discovery of
a not too old isolated NS slightly colder than the
Vela pulsar would contradict scenario ({\it 4})].  

Thus, the pycnonuclear burning 
of highly exotic atomic nuclei in accreting NSs
is helpful to solve the longstanding problem of the EOS of supranuclear
matter in NS cores. Hopefully, new observations of SXTs
and cooling isolated NSs will appear in the near future.
Combined together, they will give
more stringent constraints on the EOS of dense matter.  

\begin{acknowledgement}
The authors are grateful to P.\ Haensel, M.E. Gusakov, A.D.\ Kaminker,
A.Y.\ Potekhin, and Yu.A.\ Shibanov for discussions,
and to M.G.\ Revnivtsev and P.\ Stykovsky for 
reprocessing the observations of SAX J1808.4--3658.
DGY is grateful to JINA and the organizers of ENAM--2004
for financial support which allowed him to attend the
conference. KL acknowledges the support of the Russian Science 
Support Foundation.
This work was supported in part by
the RBRF (grants Nos. 02-02-17668 and 03-07-90200) and
the RLSS (grant 1115.2003.2).
\end{acknowledgement}

\end{document}